\documentclass{elsart}
\usepackage{amssymb}
\usepackage{graphicx}

\begin{document}
\begin{frontmatter}
\title{Distribution of repetitions of ancestors in   genealogical trees}
\author[1]{Bernard Derrida,}
\author[2]{Susanna C. Manrubia,} 
\and
\author[3]{Dami\'an H. Zanette}
\address[1]{Laboratoire de Physique Statistique de l'\'Ecole Normale
Sup\'erieure. 24 rue Lhomond, F-75231 Paris 05 Cedex, France}
\address[2]{Fritz-Haber-Institut der Max-Planck-Gesellschaft. 
Faradayweg 4-6, 14195 Berlin, Germany}
\address[3]{Consejo  Nacional de  Investigaciones  Cient\'{\i}ficas  y
T\'ecnicas. Centro At\'omico Bariloche  e  Instituto  Balseiro,
8400  S.C.  de Bariloche, R\'{\i}o Negro, Argentina}

\begin{abstract}
We calculate the  probability distribution of repetitions of ancestors
in a  genealogical   tree  for  simple neutral   models   of a  closed
population    with    sexual    reproduction    and    non-overlapping
generations. Each ancestor at generation $g$ in the  past has a weight
$w$ which is (up to a normalization) the number of times this ancestor
appears in the genealogical tree of an individual
at present. The  distribution   $P_g(w)$ of these weights   reaches  a
stationary shape  $P_\infty(w)$,   for large  $g$, i.e.   for  a large
number of generations back in the past.
For small $w$,   $P_\infty(w)$  is a   power law ($P_\infty(w)    \sim
w^{\beta}$), with a non-trivial exponent $\beta$ which can be computed
exactly using a
standard   procedure  of  the  renormalization  group   approach. Some
extensions of the model are discussed
and the  effect of these variants   on the shape of  $P_\infty(w)$ are
analysed.
\end{abstract}

\begin{keyword}
Genealogy, critical phenomena, renormalization group.
\end{keyword}
\end{frontmatter}

\section{Introduction}
Non-trivial power laws are  known  to characterize second order  phase
transitions.  A great success of the theory  of critical phenomena has
been   to  develop  methods   allowing to  predict   these power  laws
\cite{domb}.  One of the most successful approaches used in the theory
of critical phenomena is the  renormalization group, which consists in
trying  to relate physical properties of  a given  system at different
values   of the  external  parameters   (like the  temperature or  the
magnetic field). In the last three  or four decades, other non-trivial
power laws \cite{mandelbrot} have been  found in all kinds of systems:
Transition to chaos by  period doubling \cite{feigenbaum}, geometrical
problems like self avoiding  walks  (which model polymers) and  random
walks \cite{duplantier}, sand    pile models and several  other   self
organised critical systems \cite{bak}, coarsening \cite{derrida}, etc.
In many cases, renormalization ideas could  be extended to predict the
exponents of these power laws.

In this   work,   we  report  recent   results  on  simple   models of
genealogical trees  \cite{DMZ}. When one looks  at the distribution of
repetitions  in a genealogical  tree (in  the  framework of the simple
models defined  below), one  observes   non trivial  power  laws.  The
exponents  of  these power  laws can  be  calculated  {\it exactly} by
writing a  relation on the generating  function of  the weights of the
ancestors (a quantity proportional to  the number of times they appear
in a genealogical tree) which has the form of a simple renormalization
transformation.  Beyond the   intrinsic  interest of these models   to
describe real genealogies, they constitute simple pedagogical examples
for  which  renormalization ideas allow the   exact  prediction of non
trivial exponents.

\section{Neutral models of genealogical trees}

\subsection{The random parent model}
Let us first consider  a simple neutral model  of a  closed population
with sexual reproduction. By definition  of the model, the  population
size at generation   $g$ in the past is    $N_g$ and each   individual
at generation $g$ has two parents chosen at random among the $N_{g+1}$
individuals  in the  previous  generation $g+1$.  Here  $g$ counts the
number  of   past generations and   so  increases as one climbs   up a
genealogical tree. 
For simplicity we
will consider either a  population of constant  size  ($N_g= N$) or  a
population size increasing  exponentially with an average number $p/2$
of  offsprings   per couple, i.e.  \makebox{$N_g   =   \left(2 \over p
\right)^g N_0$} as $g$ counts the number of past generations; $N_0$ is
the size of  the population at present, while  the constant size  case
corresponds to $p=2$.

A related model was introduced to study the genetic
similarity between individuals in  a population evolving  under sexual
reproduction  \cite{Serva},  although  there   the  two  parents  were
distinct. We do not exclude this case here.

Clearly, the  number of branches   of the  genealogical tree  of   any
individual increases like $2^g$ and, as soon as the number of branches
exceeds  $N_g$,  there  should be  repetitions in   this tree.  Let us
denote  by $r_i^{(\alpha)}(g)$ the number of  times that an individual
$i$ living  at generation $g$ in  the past appears in the genealogical
tree of individual $\alpha$.  At generation $g=0$, the only individual
in the tree of $\alpha$ is $\alpha$ itself, therefore
\begin{equation}
r_i^{(\alpha)}(0) = \delta_{i,\alpha} 
\label{initial}
\end{equation}
and the evolution of these repetitions satisfies the recursion
\begin{equation}
r_i^{(\alpha)}(g+1) =  \sum_{j \ {\rm children \  of } \  i} 
r_j^{(\alpha)} (g).
\label{evolution}
\end{equation}

The quantity  we want to consider  is the probability $H(r,g)$ that an
individual  living at generation $g$ in  the past appears $r$ times in
the genealogical tree  of  individual $\alpha$  (living  at generation
$0$).  Normalization implies
\begin{equation}
\sum_{r \geq 0} H(r,g) = 1   \  ,
\label{norm1}
\end{equation}
the initial condition (\ref{initial}) gives
\begin{equation}
H(r,0)= {1 \over N_0} \delta_{r,1} + 
\left( 1 - {1 \over N_0} \right) \delta_{r,0} \    ,
\label{initialb}
\end{equation}
and the fact that each individual has 
two parents at the previous generation gives
\begin{equation}
\sum_{r \geq 0}  r H(r,g) = {2^g \over N_g}  .
\label{norm2}
\end{equation}

\begin{figure}[tbp]
\centerline{\includegraphics[width=9cm,angle=270]{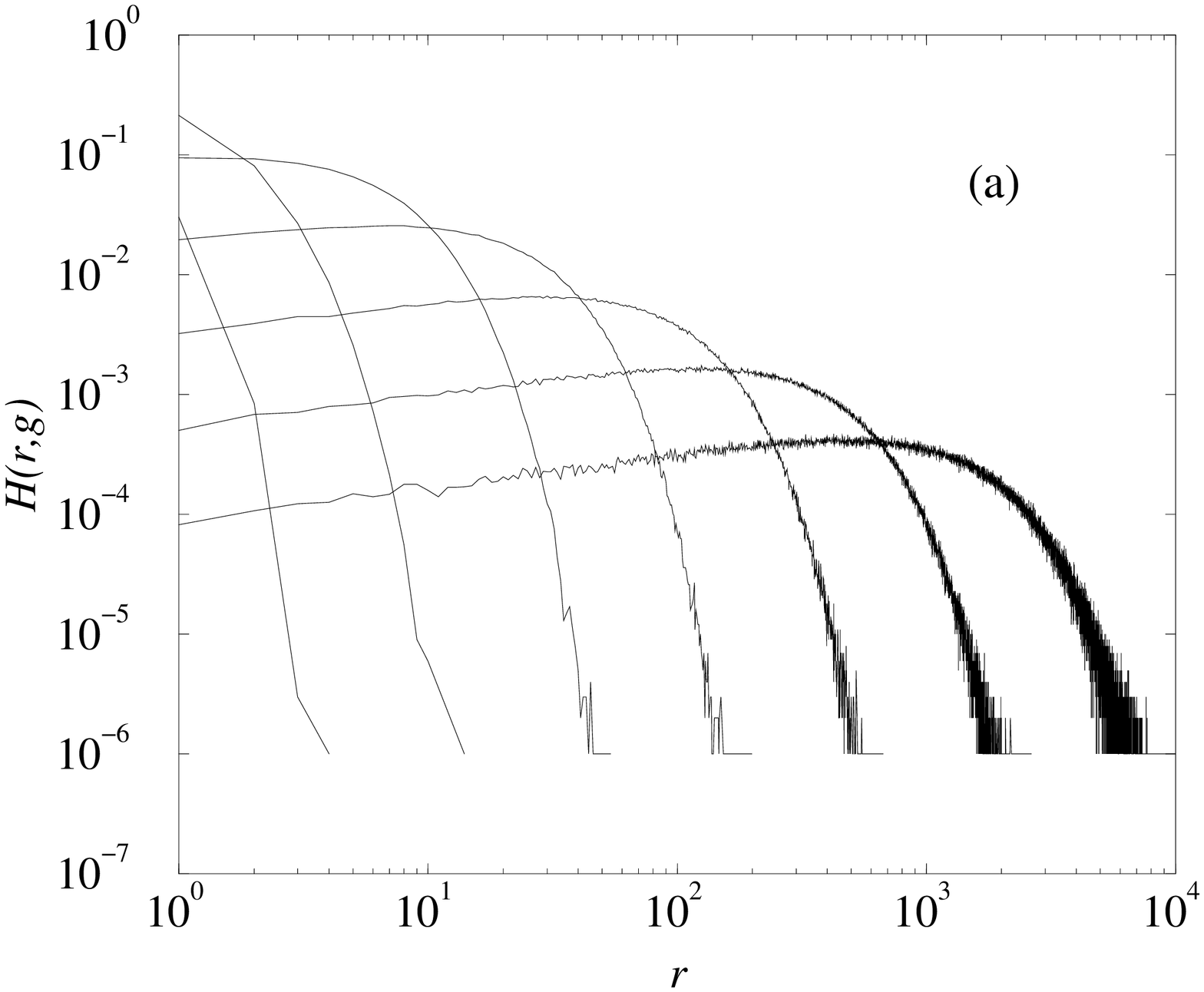}}
\centerline{\includegraphics[width=9cm,angle=270]{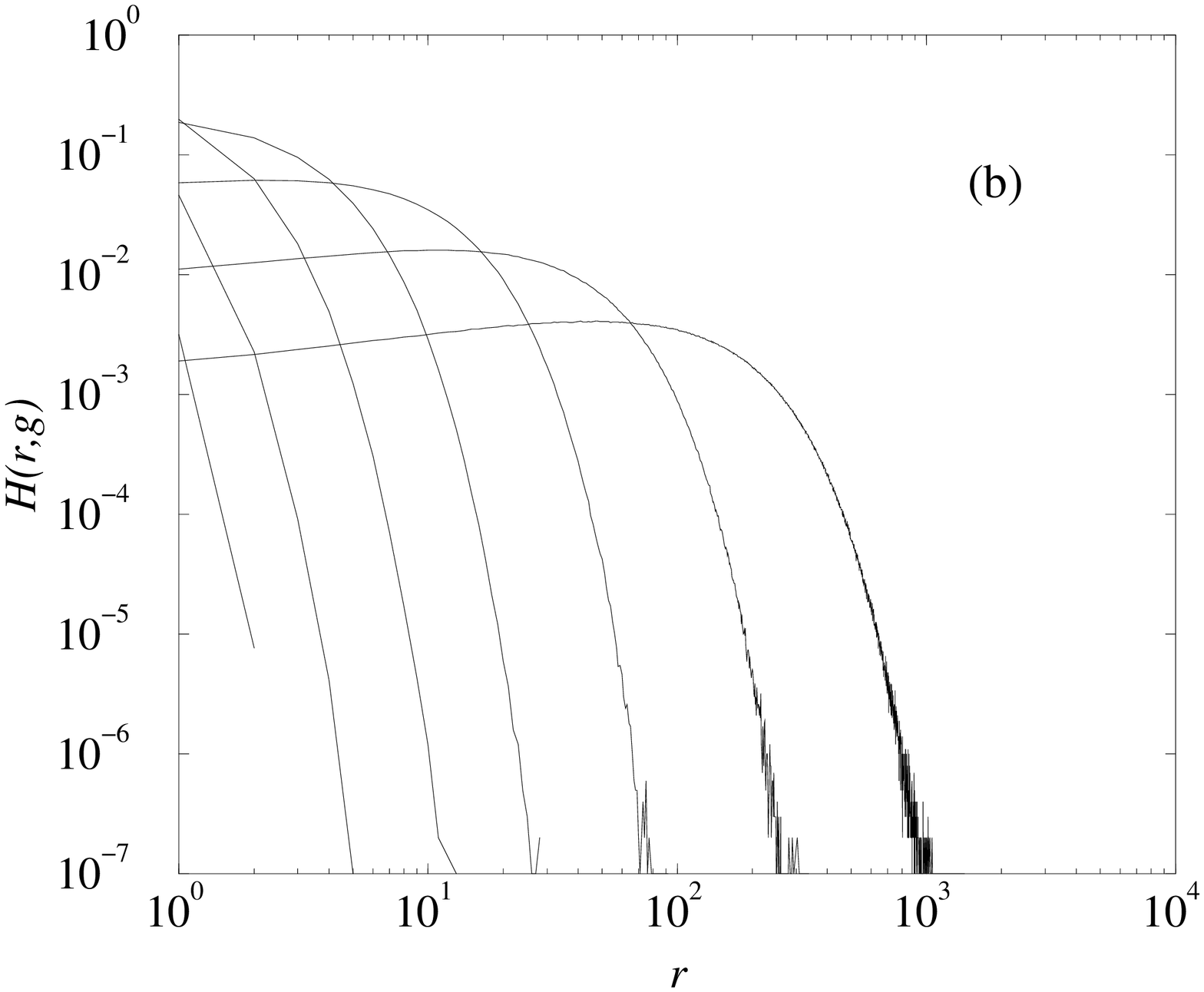}}
\caption{Probability distribution
$H(r,g)$ of $r$ repetitions after $g$ generations
($H(0,g)$ is not shown) at  $g= 5, 9, 12, 14, 16, 18,$ and $20$
for a population of constant size. In figure 1a, $N=1000$ and 
in figure 1b, $N=10000$. Both figures show averages over 1000 samples.}
\label{fig:1}
\end{figure}

These probabilities    $H(r,g)$ can be  measured  by  simulating small
systems  through a  Monte Carlo procedure:  For each  individual  of a
population  at generation $g$, two  parents are chosen at random among
the  $N_{g+1}$ individuals at  generation  $g+1$.  Figure  1 shows the
results  of such simulations  for two  populations  of constant sizes,
$N_g=N_0$ for several  values of $g$ with $N_0=  1000$ in  fig. 1a and
$N_0= 10000$ in fig. 1b.

We see that for small $g$ there  are very few repetitions and $H(r,g)$
decreases very  fast with $r$. On the  other hand, when $g$ increases,
the shape of $H(r,g)$ becomes independent of $g$ and of the population
size $N$,  with a clear  power law at  small $r$ and  a  fast decay at
large $r$.  Figure 2 
shows   the distribution  $H(r,g)$ for   several values of   $g$ and a
population which increases exponentially   with time, $N_g =  3^{10-g}
2^g$. Here, again, the shape becomes  stationary in the interval where
$g$ is large  enough and $N_g$ is still  large.  This stationary shape
is different from the one seen in fig. 1.

\begin{figure}[tbp]
\centerline{\includegraphics[width=9cm,angle=270]{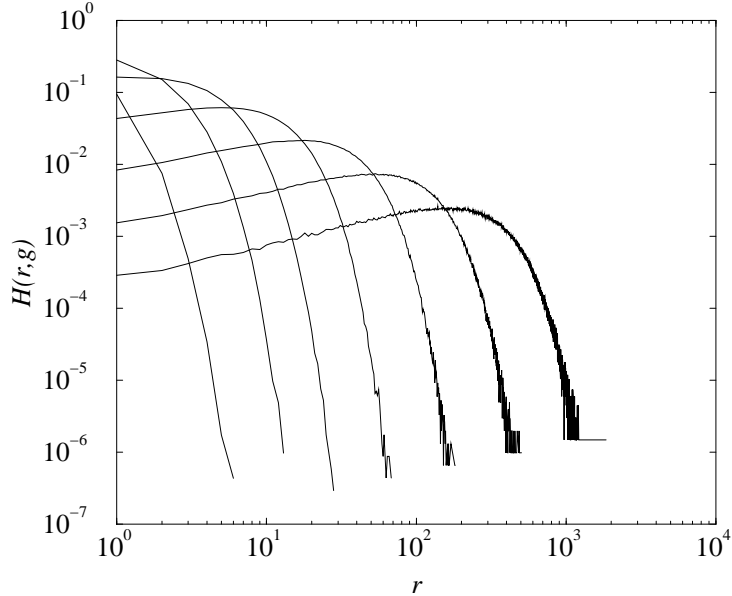}}
\caption{Probability distribution $H(r,g)$ for a population size increasing 
by a factor $3/2$ at each generation. Here $N_g = 3^{10-g} 2^g$, and 
averages over $5000$ samples are performed. The generations shown are $g=8, 
10, 11, 12, 13, 14,$ and $15$.}
\label{fig:2}
\end{figure}

The shape of $H(r,g)$ becomes stationary for large $N_g$ and large $g$
in the sense that
one gets a fixed distribution  by an  appropriate rescaling.  
In fact, introducing the rescaled quantities $w$ and $P_g(w)$
\begin{equation}
w= {N_g \over 2^g} \ r \label{wdef} 
\end{equation}
\begin{equation}
P_g(w)= {2^g \over N_g} H(r,g), \label{Pwdef} 
\end{equation}
where $w$ can be considered as a continuous variable for $N_g \ll 2^g$,   
(\ref{norm1},\ref{norm2}) transform into
\begin{equation}
\int P_g(w) dw = \int w P_g(w) dw =1,
\label{norm}
\end{equation}
and we expect $P_g(w)$  to become a fixed  distribution $P_\infty(w)$.
This means that if we associate to each  individual $i$ in the tree of
$\alpha$ at generation $g$ in the past a weight defined by
\begin{equation}
w_i^{(\alpha)} = {N_g \over 2^g} \ r_i^{(\alpha)} 
\label{wi}
\end{equation}
the distribution of these weights becomes stationary in the scaling 
limit.

  From (\ref{evolution},\ref{wi}) it is clear that these weights satisfy
\begin{equation}
w_i^{(\alpha)}(g+1) = {N_{g+1} \over 2 N_g}   \sum_{j \ {\rm children 
\  of } \  i} w_j^{(\alpha)} (g)  .
\label{evolutionw}
\end{equation}
As  we  limit ourselves  to  the   case  of  a population   increasing
exponentially at rate  $p/2$ per generation (so that  $N_g  = \left( 2
\over p \right)^g N_0 $),
(\ref{evolutionw}) reduces to
\begin{equation}
w_i^{(\alpha)}(g+1) = {1 \over p}   \sum_{j \ {\rm children \  of } 
\  i} w_j^{(\alpha)} (g)  .
\label{wp}
\end{equation}

The  ratios  $w_i^{(\alpha)}(g)/N_g$   can   be  interpreted   as  the
probability of reaching  individual  $i$ by randomly climbing  up  the
genealogical tree of $\alpha$.  In the particular case of a population
of constant size ($p=2$), the factor  $1/2$ in (\ref{wp}) is easy to
understand. For a  population of increasing size  ($p >2$), there is a
factor $1/p$ in (\ref{wp})  instead of $1/2$ because of  the factor 
$N_g$ in the definition (\ref{wi}) of the weights $w_i^{(\alpha)}$.

The key observation which   allows one to calculate  the  distribution
$P_g(w)$ in the scaling limit (large $g$ and large $N_g$) is that, for
large   $N_g$  and  for    large $g$,   {\it   the   random  variables
$w_j^{(\alpha)}$ which   appear  in the   r.h.s. of (\ref{wp})  become
independent}.  This is due to the fact that (at least  in the model we
consider) the weights $w_j^{(\alpha)}(g)$ (of brothers and sisters) in
the r.h.s. of (\ref{wp}) are uncorrelated. This independence, which is
discussed in the appendix, will be the basis of the calculation of the
fixed distribution $P_\infty(w)$ in the following sections.

\subsection{Variants of the  model}
One  can  consider  some  variants  of the  model   defined above, for
instance:
\begin{itemize}
\item
At each generation one could form fixed couples by making random pairs
and assign to each individual at generation $g$ one of these pairs (of
parents) chosen at random at  the previous generation ($g+1$). In this
case
the correlations between the weights $w_g$ would again be small in
the scaling limit and they can be ignored  in the r.h.s. of
(\ref{wp}).
\item
One can also consider an imaginary 
situation where each  individual has $p' \neq  2$ parents (instead  of
$p'=2$).   In this  case,  the  definition  of the  weights (\ref{wi})
should be replaced by
\begin{equation}
w_i^{(\alpha)} = {N_g \over (p')^g} \ r_i^{(\alpha)} 
\label{wip}
\end{equation}
to keep $P_g(w)$ normalized as in (\ref{norm}).
For a population of constant size $N_g=N$,
the evolution of the weights (\ref{wp}) becomes
\begin{equation}
w_i^{(\alpha)}(g+1) = {1 \over p'}   \sum_{j \ {\rm children \  of } 
\  i} w_j^{(\alpha)} (g).
\label{wpp}
\end{equation}
As  shown in the appendix, in  the scaling  limit, the correlations on
the r.h.s.  of (\ref{wpp}) can be neglected in this case too.

\end{itemize}
In the remaining of this work, we  try to predict the stationary shape
$P_\infty(w)$.

\section{Generating function}
The fact that the weights in the r.h.s. of (\ref{wp}) are uncorrelated
greatly simplifies   the   problem.   One  can  then   consider   that
$w_i^{(\alpha)}(g+1)$ is    the  sum of  $k$   independent identically
distributed random  variables $w_j^{(\alpha)}(g)$, where $k$ is itself
random. The probability $q_k$ of $k$ is clearly
$$q_k = {2 N_g \choose k} 
\left( 1 \over N_{g+1} \right)^k
\left( 1 - {1 \over N_{g+1}} \right)^{2 N_g -k} $$
which for large $N_g$ becomes (using the fact that $N_{g+1}  = 2 N_g /
p $) a Poisson distribution
\begin{equation}
q_k = {p^k \over k!} e^{-p} \; .
\label{poisson}
\end{equation}
Therefore for large $N_g$, the number $k$  of terms ($k$ is the number
of  children of  $i$)   in the   r.h.s.   of (\ref{wp})   is  randomly
distributed   according to (\ref{poisson})   and these  $k$ terms  are
uncorrelated.    This  becomes  a   problem   of  branching  processes
\cite{Harris}.   If      one  introduces  the    generating   function
$Q(\lambda,g)$
\begin{equation}
Q(\lambda,g) = \langle \exp[ \lambda \  w_i^{(\alpha)} (g) ] \rangle
\label{Qdef}
\end{equation}
and uses (\ref{wp}) and the fact that the weights are independent, one
finds that $Q(\lambda,g)$ satisfies
\begin{equation}
Q(\lambda,g+1) =  \sum_{k \geq 0} q_k \  Q\left( {\lambda \over p}, 
g \right)^k 
= \exp \left[-p+ p \ Q\left( {\lambda \over p}, g \right) \right].
\label{Qrecursion}
\end{equation}
The normalization (\ref{wi}) of the $w_i^{(\alpha)}(g) $ 
implies that we have for all $g$
\begin{equation}
Q(0,g) =   Q'(0,g) =1.
\label{norm5}
\end{equation}
Recursions similar  to (\ref{Qrecursion})  appear   in the  theory  of
branching  processes, in particular    in the Galton-Watson   process,
already introduced  in the 19th  century to  study the problem  of the
extinction of families \cite{Harris}.

  From (\ref{Qdef},\ref{Qrecursion}), one can easily obtain recursions 
for the moments of the weigths $w_i^{(\alpha)}$,
\begin{eqnarray}
&&\langle w(g+1) \rangle =  \langle w(g) \rangle  = 1
\label{w1rec} \\
&&\langle w^2(g+1) \rangle = {1 \over p} \langle w^2(g) \rangle  + 1
\label{w2rec} \\
&&\langle w^3(g+1) \rangle = {1 \over p^2} \langle w^3(g) \rangle + 
{3 \over p} \langle w^2(g) \rangle  + 1
\label{w3rec} \\
&&\langle w^4(g+1) \rangle = {1 \over p^3} \langle w^4(g) \rangle+ 
{4 \over p^2} \langle w^3(g) \rangle + {3 \over p^2} \langle w^2(g) 
\rangle^2 + {6 \over p} \langle w^2(g) \rangle  + 1 
\label{w4rec}
\end{eqnarray}
and  so  on.    We  see  that  for    large   $g$,   each   moment  of
$w_i^{(\alpha)}(g)$  has  a limiting  value,    as expected from   the
observation in the previous section that $P_g(w)$ converges to a fixed
distribution $P_\infty(w)$ such that
\begin{equation}
Q(\lambda,\infty) = \int_0^\infty e^{\lambda w} \ P_\infty(w) \ dw .
\label{new}
\end{equation}
The  limiting values of these moments
\begin{eqnarray}
&&\langle w^2(\infty) \rangle = {p \over (p-1)} 
\label{w2lim} \\
&&\langle w^3(\infty) \rangle =  {p^2 (p+2) \over  (p- 1)(p^2-1)}
\label{w3lim} \\
&&\langle w^4(\infty) \rangle = 
{p^3 (p^3+5p^2+6p+6) \over  (p-1)(p^2-1) (p^3-1) }
\label{w4lim}
\end{eqnarray}
etc., can  be    obtained   directly   by  expanding   the    solution
$Q(\lambda,\infty)$ of
\begin{equation}
Q(\lambda,\infty) =   \exp \left[-p+ p \ Q\left( {\lambda \over p}, \infty 
\right) \right]
\label{fixedQ}
\end{equation}
around $\lambda =0$ (choosing as normalization  $Q'(\lambda,\infty)=1$),
\begin{eqnarray}
Q(\lambda,\infty) = 1 + \lambda + {p \over 2(p-1)} \lambda^2 + {p^2 (p+2) 
\over 6 (p-1)(p^2-1)} \lambda^3 \nonumber 
\; \; \; \; \; \; \; \; \; \; \; \; \\
\; \; \; \; \; \; \; \; \; \; \; \; \; \; \; \; \; 
\; \; \; \; \; \; \; \; \; \; \; \; \; \; \; \; \; \; \; 
+ {p^3 (p^3+5p^2+6p+6) \over 24 (p-1)(p^2-1) (p^3-1) }\lambda^4 + 
 O(\lambda^5).
\label{expan}
\end{eqnarray}

Several  other properties of  $Q(\lambda,\infty)$ can be obtained from
the fixed point      equation (\ref{fixedQ}) or from  the    recursion
(\ref{Qrecursion}).  The 
simplest one is the limit
\begin{equation}
 S = \lim_{\lambda \to -\infty} Q(\lambda,\infty)  ,
\end{equation}
where $S$ is the solution ($S \neq 1$) of 
\begin{equation}
S= e^{-p + p S} 
\label{Seq}
\end{equation}
This limiting  value  ($S = 0.20318787 \dots$  for a  population of
constant  size,  i.e. $p=2$)   is  the coefficient  of  $\delta(w)$ in
$P_\infty(w)$  and so    is  the  fraction  of the   population  whose
descendants  become  extinct:  There  is a   fraction  $e^{-p}$ of the
population with no children, a fraction  \makebox{ $e^{-p+ p e^{-p}} -
e^{-p}$} of the population with  children but no grandchildren, and so
on, and the sum of all these contributions gives $S$.

Equations (\ref{Qrecursion},\ref{fixedQ}) have    the form of   a real
space  renormalization   \cite{burkhardt}. As a  consequence,  one can
predict that for   $\lambda   \to  - \infty$,   $Q(\lambda,   \infty)$
approaches its limit as a power law,
\begin{equation}
Q(\lambda, \infty) - S \sim   |\lambda|^{-\beta -1},
\label{power}
\end{equation}
where the exponent $ \beta$  must be
\begin{equation}
\beta  = -2 - {\ln S \over \ln p}
\label{betaeq}
\end{equation}
for the  terms of order $ |\lambda|^{-\beta -1} $  on both sides of
(\ref{fixedQ}) to be 
equal.  For $p=2$, this    gives  $\beta = 0.2991138 \dots$   and
(\ref{new}) implies that  at small $w$, the distribution $P_\infty(w)$
is a power law
\begin{equation}
P_\infty(w)  \sim   w^{\beta }
\label{powerlaw}
\end{equation}
with $\beta$ given   by (\ref{betaeq}),    in agreement   with  the
results of the simulations shown in figures 1 and 2.

In fact, for $\lambda \to  - \infty$, the  leading contribution in the
difference $Q(\lambda, \infty) - S$ consistent with (\ref{fixedQ}) 
is
\begin{equation}
Q(\lambda, \infty) - S \simeq    |\lambda|^{-\beta -1} \ 
F_p \left(\ln \lambda \over \ln p \right)
\label{oscillations}
\end{equation}
where $F_p(z)$  is   an arbitrary periodic  function  (not necessarily
constant) of period $1$  (i.e.   $F_p(z+1) = F_p(z)$).  Such  periodic
amplitudes are often present in the critical behavior of systems which
have a discrete  scale invariance \cite{DGP}.  It is easy
to calculate numerically the    function $Q(\lambda,\infty)$ for   all
values   of  $\lambda$ from  the  fixed   point equation (\ref{fixedQ})
 which relates
$\lambda$ to  points  $\lambda /  p^n$  
arbitrarily    close     to $0$,  where    the    linear approximation
$Q(\lambda,\infty)  \simeq  1  +  \lambda    = O(\lambda)^2$   becomes
excellent.  Using  this procedure, we could  determine  (figure 3) the
combination $[Q(\lambda,\infty) -  S] | \lambda|^{-\beta -1}$ and  the
non constant  periodic nature of the  amplitude $F_p(z)$ is visible if
$p$ is  large  enough.  The analytic  determination  of $F_p(z)$ is in
principle possible \cite{hakim,hm} for $p $ close  to $1$, but remains
difficult for arbitrary $p$.

\begin{figure}[tbp]
\centerline{\includegraphics[width=9cm,angle=270]{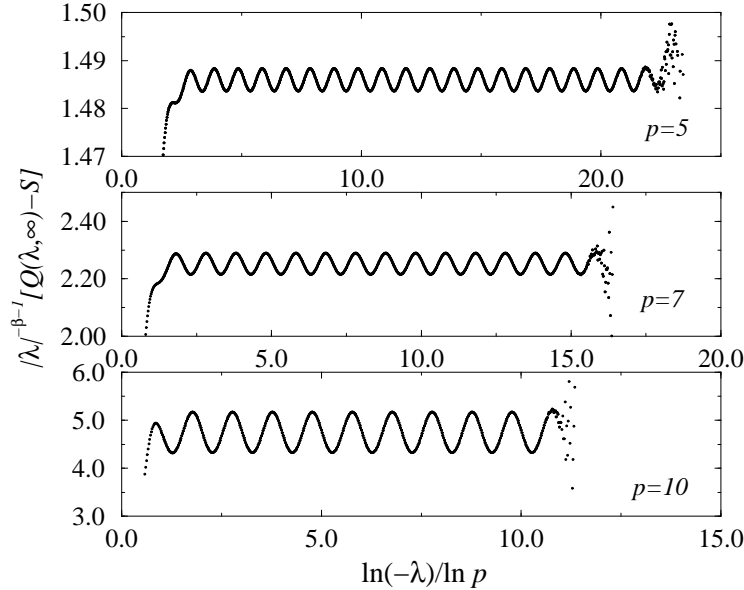}}
\caption{The product $|\lambda|^{-\beta-1} [ Q(\lambda,\infty) - S]
$ versus $\ln (- \lambda) / \ln p$  for $p=5,7,10$. We see clearly
the periodic nature of the amplitude predicted by (\ref{oscillations}).
Discrepancies at small $-\lambda$ are due to the fact that the asymptotic 
regime is not yet reached. At too large $-\lambda$, rounding errors in the 
difference $Q(\lambda,\infty) - S$ make the resulst noisy and unreliable.}
\label{fig:3}
\end{figure}

The   knowledge of  the  periodic   function  $F_p(z)$  determines  in
principle  the whole  expansion of  $Q(\lambda,  \infty)$ in the limit
$\lambda \to -  \infty$.  If we look for  a solution of (\ref{fixedQ})
which starts as (\ref{oscillations})  as $\lambda  \to - \infty$,  one
finds by equating  the two sides of (\ref{fixedQ})  order  by order in
powers of $|\lambda |^{-\beta -1}$,
\begin{eqnarray}
Q(\lambda,\infty)= S + {F_p \left(\ln \lambda \over \ln p \right)   
\over |\lambda|^{\beta + 1}} + {p\over 2 (pS-1) }
\left[{ F_p \left(\ln \lambda \over \ln p \right) \over 
|\lambda|^{ \beta +1}} \right]^2 \nonumber 
\; \; \; \; \; \; \; \; \; \; \; \; \\
\; \; \; \; \; \; \; \; \; \; \; \; \; \; \; \; \; 
\; \; \; \; \; \; \; \; \; \; \; \; \; \; \; \; \; \; \; 
+ {p^2 (pS+2) \over 6 (pS- 1)((pS)^2-1)}
\left[{F_p \left(\ln \lambda \over \ln p \right)
 \over |\lambda|^{ \beta + 1}}  \right]^3+ \dots
\label{expan7}
\end{eqnarray}

In addition  to the moments (\ref{w2lim}-\ref{w4lim}) of $P_\infty(w)$
(which    are    given   by     the   expansion   (\ref{expan})     of
$Q(\lambda,\infty)$)  and the exact values (\ref{Seq},\ref{betaeq}) of
$S$ and $\beta$, let us just mention two properties of the solution of
(\ref{fixedQ}) which we checked  
by rather involved ways, and that we prefer to leave as conjectures:

\begin{itemize}
\item $Q(\lambda,\infty)$ is analytic  in  the whole complex plane  of
$\lambda$

\item $Q(\lambda,\infty)$ grows 
extremely fast (faster than the exponential of  the exponential ... of
the   exponential  of $\lambda$)   as  $\lambda  \to   \infty$.  As  a
consequence, for large  $w,  \  P_\infty(w)$  decays faster than   any
exponential but  slower than any   stretched exponential (of  exponent
larger than 1) and even
\begin{equation}
 1 \ll { - \ln P_\infty(w) \over w} \ll \ln w .
\label{largew}
\end{equation}
\end{itemize}

All the discussions of the present section can be repeated in the case
of  having $p'$ parents.  If  we limit ourselves   to a population  of
constant size  (as  we  did  to obtain   (\ref{wpp})),  we find   that
$Q(\lambda,\infty)$  satisfies  the     same   fixed  point   equation
(\ref{fixedQ}) as above with $p$ replaced by $p'$
\begin{equation}
Q(\lambda,\infty) =   \exp \left[-p'+ p' Q\left( {\lambda \over p'}, 
\infty \right) \right].
\end{equation}
This means  that  the distribution of the  weights  $w$ is exactly the
same for the cases of (i) 2 parents and a population size increasing
exponentially  by a factor $p/2$  at  each generation and  (ii) a
population of constant size with $p$ parents per individual.  This can
be checked   by comparing figure  2  
and figure 4, where we show the distributions $H(r,g)$ for a population
of  constant  size $N =  1000$   and $N=10000$  with $3$  parents  per
individual.

\begin{figure}[tbp]
\centerline{\includegraphics[width=9cm,angle=270]{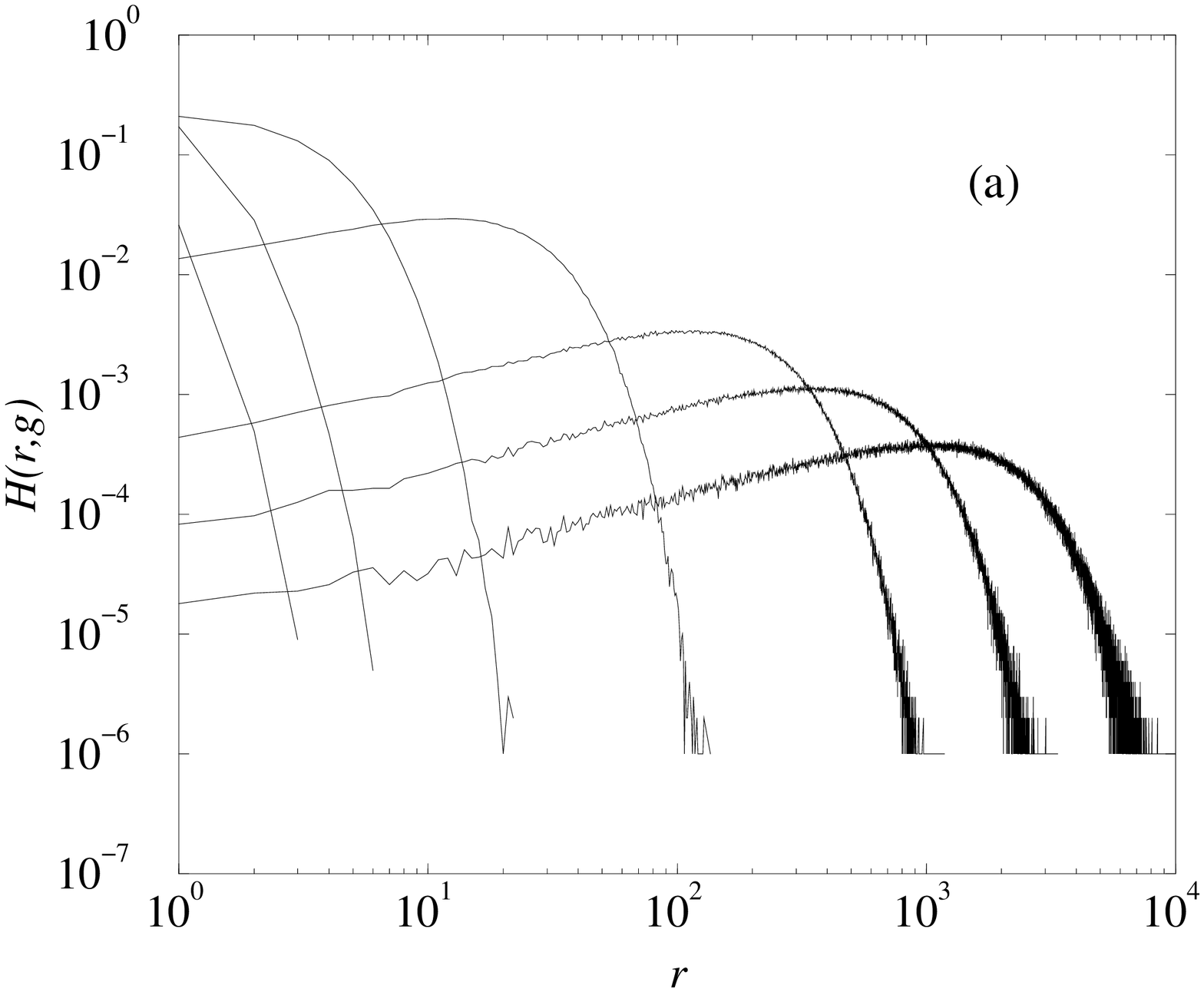}}
\centerline{\includegraphics[width=9cm,angle=270]{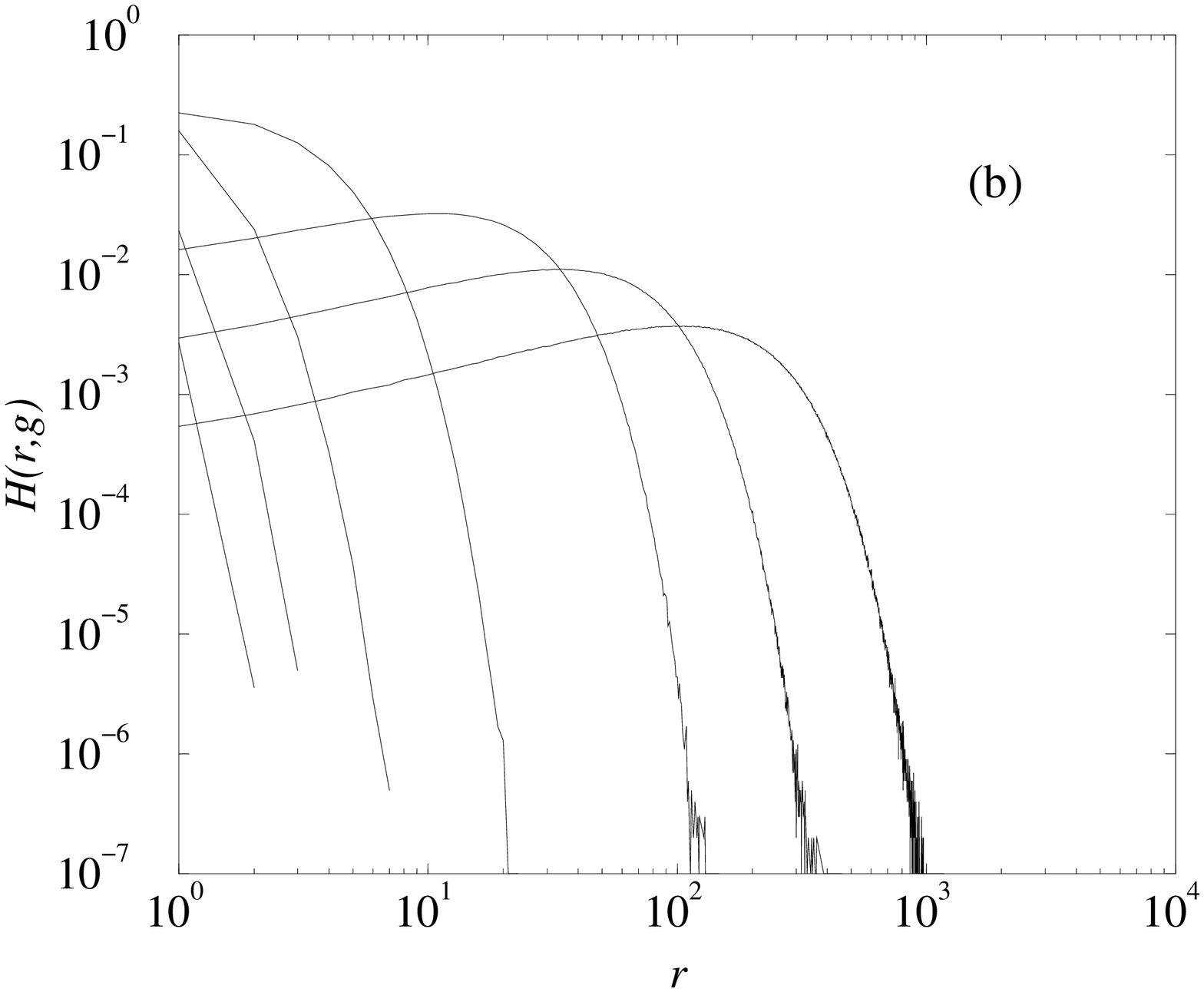}}
\caption{The function $H(r,g)$ for a population of constant size
with (a) $N=1000$ and (b) $N=10000$ when the number $p$ of parents is
3. The generations shown are $g=3, 5, 7, 9, 11, 12,$ and $13$.}
\label{fig:4}
\end{figure}

\section{Perturbation theories}
Despite its simplicity, it is not  easy to extract more information on
the  function  $Q(\lambda,\infty)$  and   consequently  on  the  
distribution      $P_\infty(w)$ from   the    fixed   point   equation
(\ref{fixedQ}). There are however two limiting  cases around which one
can 
apply a  perturbation theory and extract  a few  more properties of the
fixed distribution: $p$ close to 1 and $p$ very large.

\subsection{ $p $ close to $1$}
One  can see from (\ref{w2lim}-\ref{w4lim}) that   when $p \to 1$, the
successive moments of the weight $w$ diverge like $\langle w^n \rangle
\sim (p-1)^{1-n}$.  This indicates that if one writes
\begin{equation}
p=1 + \epsilon
\label{epsdef}
\end{equation}
the solution of the fixed point equation (\ref{fixedQ}) can be expanded
in the following way
\begin{equation}
Q(\lambda,\infty)= 1 +\epsilon f_1\left({\lambda\over \epsilon} \right)
+ \epsilon^2  f_2\left( {\lambda \over \epsilon} \right)
+ \epsilon^3  f_3\left( {\lambda \over \epsilon} \right)
+ \epsilon^4  f_4\left( {\lambda \over \epsilon} \right) +... 
\label{expanbis}
\end{equation}
where the functions $f_1, f_2, ...$ resum the most divergent terms in 
the 
perturbative  expansion   (\ref{expan})   in  the   range $\lambda   =
O(\epsilon)$.    If we  insert   the  expansion (\ref{expanbis})  into
(\ref{fixedQ})  we get, by equating  the two sides   order by order in
$\epsilon$, a hierarchy of   differential equations for the  functions
$f_1, f_2, ...$ which can be solved and lead to
\begin{eqnarray}
&& f_1 (y)= { y \over 1 - {y \over 2}}  \label{f1} \\
&& f_2 (y)= {2 \over 3} { y^2 \over \left(1 - {y \over 2} \right)^2}
+{1 \over 3} { y \over \left(1 - {y \over 2} \right)^2}  \ln \left[ 1 - 
{y \over 2} \right] \label{f2}  \\
&& f_3 (y)= {14 y^3 - 3 y^2 \over 36  \left(1 - {y \over 2} \right)^3}
+ {17 y^2 - 6 y \over 36  \left(1 - {y \over 2} \right)^3}  \ln \left[ 
1 - {y \over 2} \right] \nonumber \\ 
&& \; \; \; \; \; \; \; \; \; \; \; \; \; \; \; \; \; \; \; \; \; \; 
\; \; \; \; \; \; \; \; \; \; \; \; \; \; \; \; \; \; \; \; \; \; \; 
\; \; \; \; \; \; \;
+ { y^2 + 2 y \over 36  \left(1 - {y \over 2} \right)^3}  \ln^2 
\left[ 1 - {y  \over 2} \right].  \label{f3} 
\end{eqnarray}
Comparing these expressions for  large negative $y$ with (\ref{expan7}), 
one gets the expansions of $S$, $\beta$ 
$$ S= 1 - 2 \epsilon + { 8 \over 3} \epsilon^2 -{28 \over 9} \epsilon^3 
+  O \left( \epsilon^4 \right) $$ 
$$ \beta = { \epsilon \over 3} - {\epsilon^2 \over 18} + {19 \over 540
}  \epsilon^3 + O  \left( \epsilon^4 \right)  $$ 
which both  agree   with what one   would  get by directly expanding
(\ref{Seq},\ref{betaeq}).  What the  small $\epsilon$ expansion  gives
us in addition   is the  function  $F_p(z)$  which  is found  to be  a
constant function of $z$ to all orders in powers of $\epsilon$,
$$ F_p(z) =  4 \epsilon^2 - {32 \over 3}  \epsilon^3 + 18 \epsilon^4 
+ O \left( \epsilon^5 \right). $$  The non-constant nature of $F_p(z)$
does not show  up in the expansion  in powers of  $\epsilon$. It  is a
non-perturbative contribution  (which  vanishes   to  all orders    in
$\epsilon= p-1$) which   could be calculated \cite{hakim}  using 
WKB-like techniques \cite{hm}.

\begin{figure}[tbp]
\centerline{\includegraphics[width=9cm,angle=270]{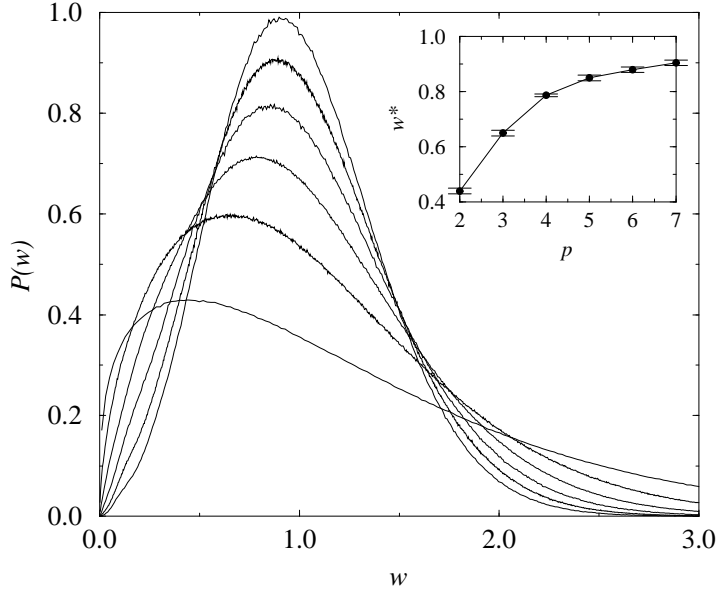}}
\caption{The  fixed distribution $P_\infty(w)$ (the delta function 
contribution at $w=0$ is not shown) for $p=2$, with $N=2^{15}$ and
$g=25$; $p=3$, $N=3^{10}$, $g=18$; $p=4$, $N=4^8$, $g=14$; $p=5$,
$N=5^6$, $g=11$; $p=6$, $N=6^6$, $g=11$; and $p=7$, $N=7^5$, $g=9$.
Averages over 1000 realizations have been carried out.
The insert shows how the maximum $w^*$ varies with $p$.}
\label{fig:5}
\end{figure}

 From  (\ref{expanbis}-\ref{f2})  and   the definition  (\ref{new}) one
finds that, for small $\epsilon$, the continuous part
of $P_\infty(w)$
is  an exponential
$$P_\infty(w) \simeq \left( 1 -2 \epsilon + {8 \epsilon^2 \over 3} 
\right) \delta(w) \ + \ 4 \epsilon^2 e^{-2 \epsilon w} . $$
Corrections to this exponential shape are extractable from higher 
order terms  ($f_2$, $f_3$, \dots ).

\subsection{ large $p$}
The other  case which can be 
dealt with perturbatively is the limit of large  $p$.  If $p$ is large
and $\lambda = O(p^{1/2})$, the solution of (\ref{fixedQ}) is given by
\begin{eqnarray}
 \ln Q(\lambda,\infty)   =  \lambda + {\lambda^2 \over 2 p } 
 +    {\lambda^3 \over 6 p^2 } \nonumber 
 +    \left[ {\lambda^2 \over 2 p^2}+ {\lambda^4 \over 24 p^3 } \right] 
\nonumber \\
\; \; \; \; \; \; \; \; \; \; \; \; \; \; \; \; \; \; \; \; \; \; \; \;
\; \; \; \; \; \; \; \; \; \; \; \; \; \; \; \; \; \; \; \; \; \; \; \;
+ \left[{\lambda^3 \over 2 p^3}+ {\lambda^5 \over 120 p^4 } \right]     
+  O(p^{-2}) \; \; ,
\end{eqnarray}
where each term represents a new order  in powers of $p^{-1/2}$.  This
implies that $P_\infty(w)$ can be written in terms of $x=w-1$ in the
range $x \sim p^{-1/2}$ as
\begin{eqnarray}
&&P_\infty(w)   \simeq  \sqrt{p \over 2 \pi} e^{-px^2/2}   \left[ 1
 +   \left( { p x^3 \over 6} \nonumber  - {x \over 2} \right)
 +   \left( {p^2 x^6 \over 72} - {p x^4 \over 6} + {7 x^2 \over 8} - 
{7 \over 12 p} \right) \right. \nonumber \\
&&+ \left. \left( {p^3 x^9 \over 1296} -{p^2 x^7 \over 48}+ 
{19 p x^5 \over 80} -{95 x^3 \over 144} -{x \over 8 p} \right)+ 
\dots \right] \; \; ,
\label{largep}
\end{eqnarray}
where  each parenthesis  represents  a new  order  in $p^{-1/2}$.  The
Gaussian shape in (\ref{largep})  is not a surprise  considering that,
for large $p$,  each  weight becomes  the  sum  of a large  number  of
independent contributions.

One property which can be  extracted from (\ref{largep}) is the 
location of the maximum $w^*$ of $P_\infty(w)$
\begin{equation}
w^* = 1 - {1 \over 2p} - {25 \over 24 p^2}  +  O \left( 1 \over 
p^3 \right).
\label{wsp}
\end{equation}
Figure 5   shows the shapes   (obtained  by random samplings
populations of constant sizes with $p$ parents per individual)   of the
distribution   $P_\infty(w)$ for several  choices  of  $p$. The insert
shows the values of  $w^*$ extracted from these  data. They agree with
the  prediction  (\ref{wsp}) that   the maximum  approaches   $1$ with
corrections of order $1/p$ as $p$ becomes large.

\section{Conclusions}

We have seen that for  simple neutral models  of evolution with random
mating, the distribution  of ancestors repetitions in the genealogical
tree of  a present individual becomes  stationary, with  a fixed shape
$P_\infty(w)$ which can be described by a  fixed point equation of the
type (\ref{fixedQ}).   This shape  is  the  same if one   considers  a
population increasing exponentially  at rate $p/2$ per generation with
two parents per individual  or a population of  constant size with $p$
parents per individual.

The  fixed   point  equation (\ref{fixedQ})   allows  one to determine
exactly  the      exponent   $\beta$  which   characterizes
$P_\infty(w)$   at small  $w$.   The  determination  of $\beta$   from
(\ref{fixedQ}) is  very reminiscent of the  way one finds exponents in
the renormalization   group  approach of   critical  phenomena.  Other
properties (large $w$ behavior, amplitude  of the power law, \dots) of
the fixed distribution $P_\infty(w)$ are in principle extractable from
(\ref{fixedQ}) but are  more  difficult  to obtain than   the exponent
$\beta$.
 
The present work admits   several extensions. In particular, one   may
consider  the case where the   probabilities $q_k$ (that an individual
has   $k$   children)  is  arbitrary (instead   of    Poissonian as in
(\ref{poisson})).  The  fixed  point equation  (\ref{fixedQ})  becomes
then simply
$$Q(\lambda,\infty) = \sum_k q_k \  Q \left( {\lambda \over p}, 
\infty \right)^k$$
and starting from  this new fixed  point equation, one can essentially
repeat all the above  calculations, including the determination of the
exponent $\beta$.  If all the $q_k$ vanish for $k  > k_{\rm max}$, one
can see that for large $\lambda$,
$$ \ln Q(\lambda,\infty)  \sim \lambda^{\ln k_{\rm max}/\ln p}. $$ 
Consequently,   the   distribution  $P_\infty(w)$ becomes  a stretched
exponential for large $w$,
$$\ln P_\infty(w)\sim -w^{\ln k_{\rm max}/\ln (k_{\rm max}/ p)}. $$

Recursions similar  to    (\ref{wp})  describe the   distribution   of
constraints in granular media \cite{Copper}. In such cases, the number
of grains in direct contact and supporting the weight of a given grain
is variable. This would correspond to considering that the number $p'$
of parents is  no longer constant over  the  whole population  but may
vary from 
individual to individual.

Finally let us  mention that an interesting aspect  of  the problem is
the calculation of the correlations between the genealogies of several
contemporary individuals.  One can show   \cite{DMZbis} that for large
$g$, the  weights of all the  ancestors of two distinct individuals in
the same population become the same after a number of generations $g_c
\propto \ln N$.

\section{Appendix: The correlations of the weights}

In  this appendix  we show, by  calculating moments  of the  weights $
w_j^{(\alpha)} (g)$, that    correlations  become negligible  in   the
r.h.s. of (\ref{wp}) and (\ref{wpp}).

\subsection{The case of a varying population size  with $2$ parents per 
individual}
It is convenient to rewrite (\ref{wp}) as
\begin{equation}
w_i^{(\alpha)}(g+1) = {1 \over p}   \sum_{j =1}^{N_g}  \chi(i,j  ) \
  w_j^{(\alpha)} (g)
\label{wp1}
\end{equation}
where
\begin{equation}
\chi(i,j)= \cases{
0 \ {\rm{ if}} \ i \ {\rm  is \  not \  a \  parent \  of} \  j \cr
1 \ {\rm if} \ i \ {\rm  is \  one \ of \ the \ two  \  parents \  of}
 \  j \cr 2 \ {\rm if} \ i \  {\rm is \   the \ two  \  parents \  of} 
\  j }.
\end{equation}

For the random parent model of section 2 (where each parent of $j$ is
chosen at   random  among   all   the individuals  of    the  previous
generation), $\chi(i,j)  = 0$ with probability  $(1  - 1 /N_{g+1})^2$,
$\chi(i,j) =  1$  with probability $2(1   - 1 /N_{g+1})  /N_{g+1}$ and
$\chi(i,j) = 2$   with probability  $ 1  /N_{g+1}^2$  (as  we did  not
exclude  choosing the  same   parent twice).    Moreover there  is  no
correlation between    $\chi(i,j)$ and   $\chi   (i',j')$   if  $j\neq
j'$.  Lastly $\chi(i,j)$ and  $\chi(i',j)$ are correlated  for $i \neq
i'$ and
\begin{equation}
\langle \chi(i,j) \chi (i',j) \rangle  = {2 \over N_{g+1}^2} \; .
\label{cor1}
\end{equation}
This correlation together with
\begin{eqnarray}
&&\langle \chi(i,j)  \rangle  = {2 \over N_{g+1} }
\label{cor2} \\
&&\langle \chi(i,j)^2  \rangle  =  {2 \over N_{g+1}} + {2 \over N_{g+1}^2} 
\label{cor3} \\
&&\langle \chi(i,j) \chi(i',j') \rangle  = {4 \over N_{g+1}^2}  \ \ \ \ \
 \ \ \ \ \ \ {\rm for} \ \    j \neq j'
\label{cor4}
\end{eqnarray}
when used in (\ref{wp1}) 
leads to
\begin{eqnarray}
\langle w_i(g+1) \rangle = \langle w_i (g) \rangle
\nonumber
\end{eqnarray}
as expected, since the  definition (\ref{wdef}) of  $w$ was chosen  to
keep $\langle w \rangle =1$, and
\begin{eqnarray}
\langle w_i(g+1)^2 \rangle =  
\left( {1 \over p} + {1 \over p N_{g+1}} \right) \langle w_i (g)^2 \rangle + 
\left({1 } - {2 \over p N_{g+1}} \right) \langle w_i(g)w_{i'}(g) \rangle
\label{w2a} \\
\langle w_i(g+1) w_{i'} (g+1) \rangle =  
 {1 \over p N_{g+1}}  \langle w_i (g)^2 \rangle + 
\left(1 - {2 \over p N_{g+1}} \right) \langle w_i(g)w_{i'}(g) \rangle
\label{wwa} 
\end{eqnarray}
where $i \neq i'$  (the index $^{(\alpha)}$ has been omitted for simplicity).

 From (\ref{initial},\ref{evolution},\ref{wdef}), we know that $ \sum_i
w_i(g)=N_g$, and $\langle w_i(g) \rangle =1$. Thus for $i \neq i'$
\begin{equation}
\langle w_i(g) w_{i'} (g) \rangle = {N_g - \langle w_i(g)^2 \rangle \over
 N_g -1} 
\label{wwb} 
\end{equation}
and (\ref{w2a}) becomes
\begin{eqnarray}
\langle w_i(g+1)^2 \rangle  =
\left( {1 \over p} - {1 \over p N_{g+1}} - {1 \over N_g -1} + 
{2 \over p N_{g+1} 
( N_g -1) } \right)
\langle w_i(g)^2 \rangle  \nonumber \\
\; \; \; \; \; \; \; \; \; \; \; \; \; \; \; \; \; \; \; \; \; \; \; \; 
 + \left( 1 - {2 \over p N_{g+1}}  \right) { N_g \over N_g -1}.
\label{w2b} 
\end{eqnarray}
So far this evolution equation is exact.

If we consider that all the $N_g$'s are very large, 
(\ref{w2b}) becomes
\begin{equation}
\langle w_i(g+1)^2 \rangle  = {1 \over p} \langle w_i(g)^2 \rangle +1 \; ,
\label{w2c} 
\end{equation}
so that  for large $g$  (in fact $g$ should  not be too large  to keep
$N_g$ large  enough, more precisely $g$  should be such  that $(p/2)^g
\ll N_0 \ll  p^g $), the second  moment of $w$  has a limiting value $
\langle   w_i(g)^2 \rangle   \to {p   \over  p-1}$ and    we  see from
(\ref{wwb}) that
\begin{equation}
\langle w_i(g) w_{i'} (g) \rangle  \to 1 = \langle w \rangle^2. 
\label{wwe} 
\end{equation}
When one repeats the above calculation for higher correlations
(we did it 
up to  three-point  correlations), one   finds  that the  correlations
between the  terms in the r.h.s. of  (\ref{wp1}) are  negligible. This
indicates that these  correlations  can   be neglected (of   course  a
complete  proof  that all correlations are   negligible in the scaling
limit would be much better than our guess  based on the computation of
the lowest correlations).

One  can   repeat the above calculation    of correlations for several
variants of the  model, like those discussed at  the end of section 2.
The  exact  formulae  (\ref{w2a},\ref{wwa},\ref{w2b}) are modified but
one  always  find that,    in   the scaling  regime,  they  reduce  to
(\ref{w2c},\ref{wwe}), meaning that the correlations could be ignored.

\subsection{The case of a population of constant size  with $p'$ 
parents  per individual}

Let us consider only the case where  each individual has $p'$ parents.
To keep the  notations simple,  we  will limit the calculation  to the
case of a population of constant size
$$N_g=N$$ 
One can then follow the same steps  as above.  Starting from
(\ref{wpp}), one replaces (\ref{wp1}) by
\begin{equation}
w_i^{(\alpha)}(g+1) = {1 \over p'}   \sum_{j =1}^{N_g}  \chi(i,j  ) \
  w_j^{(\alpha)} (g).
\label{wp5} 
\end{equation}
The correlations (\ref{cor1}-\ref{cor4})  become in this case
\begin{eqnarray}
&& \langle \chi(i,j) \chi (i',j) \rangle  = {p'(p'-1) \over N^2}
 \ \ \ \ \ \ \ \ \ \ \ {\rm for} \ \    i \neq i'
\label{cor1a} \\
&&\langle \chi(i,j)  \rangle  = {p' \over N }
\label{cor2a} \\
&&\langle \chi(i,j)^2  \rangle  =  {p' \over N} + {p'(p'-1) \over N^2} 
\label{cor3a} \\
&&\langle \chi(i,j)\chi(i',j') \rangle ={p'^2 \over N^2}  \ \ \ \ \ \ 
\ \ \ \ \ {\rm for} \ \    j \neq j'
\label{cor4a}
\end{eqnarray}
and (\ref{w2a},\ref{wwa}) read 
\begin{eqnarray}
\langle w_i(g+1)^2 \rangle =  
\left( {1 \over p'} + {p'-1 \over p' N} \right) \langle w_i (g)^2 \rangle + 
\left({1 } - {1 \over  N} \right) \langle w_i(g)w_{i'}(g) \rangle
\label{w2d} \\
\langle w_i(g+1) w_{i'} (g+1) \rangle =  
 {p'-1 \over p' N}  \langle w_i (g)^2 \rangle + 
\left(1 - {1 \over  N} \right) \langle w_i(g)w_{i'}(g) \rangle.
\label{wwd} 
\end{eqnarray}

For large  $g$  and large  $N$, we  see (using the   fact that $\sum_i
w_i(g)=N$) that $\langle w_i(g)^2 \rangle \to p'/(p'-1)$ and  $\langle
w_i(g) w_{i'}(g)\rangle \to 1$ as $g \to \infty$. This again indicates
that correlations can be neglected for large $g$ and large $N$.

\section*{Acknowledgements}

Interesting discussions with Ugo Bastolla and Vincent Hakim are 
gratefully acknowledged. SCM acknowledges support from the Alexander 
von Humboldt Foundation (Germany).

\end{document}